\documentclass[aps,pra,preprint,groupedaddress,showpacs]{revtex4}

\begin{document}


\title{Dynamical decoherence in a cavity with a large number of two-level atoms}


\author{Marco Frasca}
\email[e-mail:]{marcofrasca@mclink.it}
\affiliation{Via Erasmo Gattamelata, 3 \\
             00176 Roma (Italy)}


\date{\today}

\begin{abstract}
We consider a large number of two-level atoms interacting with the mode of
a cavity in the rotating-wave approximation (Tavis-Cummings model). We apply the
Holstein-Primakoff transformation to study the model in the limit of the number
of two-level atoms, all in their ground state, becoming very large. The unitary
evolution that we obtain in this approximation is applied to a macroscopic
superposition state showing that, when the coherent states forming the superposition
are enough distant, then the state collapses on a single coherent
state describing a classical radiation mode. This appear as a true dynamical
effect that could be observed in experiments with cavities.
\end{abstract}


\maketitle


\section{Introduction}

The foundations of quantum optics rely on the Hamiltonian of the interaction of
a single radiation mode with an atom that, for all practical applications in
optical regime, reduces to the well-known Jaynes-Cummings model \cite{sch}.
The proper working of this model is due to the effectiveness of certain
approximations \cite{fra0}: The rotating wave approximation and the two-level
approximation. These approximations are enough to justify the success of the
Jaynes-Cummings model in the optical regime.

In experiments involving such a model a quite common effect that appears is
decoherence, that is the loss of unitarity in the quantum evolution due to
external environmental effects \cite{zur}. Decoherence can make very difficult
to realize quantum computers. Indeed, our aim in this paper is to treat
the problem of quantum coherence for a system obeying the Jaynes-Cummings
model and being composed by N two-level atoms in the limit of N becoming
infinitely large. Such a model is known in the literature as the Tavis-Cummings
model \cite{tc}.

Recent works on decoherence \cite{boni} have enlarged the meaning of this mechanism to
a general reduction of the density matrix to the mixed form removing interference
without any need of coupling to an external environment. The possible existence
of such kind of decoherence without environment is, indeed, an open question.

Decoherence can also be seen as an intrinsic effect due to the unitary evolution
when the number of particles becomes very large \cite{fra1,fra2,fra3}.
A first example in this sense has been given by Gea-Banacloche \cite{gb} that computed asymptotic
states for the radiation field in the Jaynes-Cummings model in
the limit of a large photon number. Such states have been recently observed in
experiments
by Haroche et al. \cite{har1,har2} proving the emergence of classical
behavior already with very few photons. The work of Gea-Banacloche appears in some
way dual to the one we present here.

In order to study the Tavis-Cummings Hamiltonian in the limit of a very large
number of particles on a macroscopic superposition state we will adopt an approach
firstly devised by Persico and Vetri \cite{pv} and recently applied in a paper
by Berman et al. \cite{ber}. These authors use the Holstein-Primakoff \cite{hp} transformation
to change the contribution of the two-level atoms into a bosonic field. In this way,
one gets the Hamiltonian of two coupled harmonic oscillators that can be easily
diagonalized. The main result of this paper will be to show how the Tavis-Cummings
Hamiltonian can produce decoherence, intrinsically in this approximation, making this effect
observable in experiments with cavities. The unitary evolution is applied to
a superposition state of two coherent states of radiation. When the distance between
such states is large enough, as already seen in
recent experiments \cite{haro, wine}, decoherence sets in and the superposition disappears
leaving just a coherent state.

The paper is structured in the following way. In sec.\ref{sec2} we introduce the
model and the Holstein-Primakoff transformation. In sec.\ref{sec2b} the
higher order corrections to the unitary evolution are evaluated. In sec.\ref{sec3} the unitary
evolution of a macroscopic superposition state is obtained proving that
decoherence is produced when the components of the superposition state are
significantly distant. In sec.\ref{sec4} the conclusions are given.

\section{Tavis-Cummings Model and Holstein-Primakoff Transformation}
\label{sec2}

The N-atom Jaynes-Cummings model was firstly considered by Cummings and
Tavis \cite{tc} and can be written as
\begin{equation}
    H = \omega a^\dagger a +\frac{\Delta}{2}\sum_{i=1}^N\sigma_{3i}+
	g\sum_{i=1}^N(\sigma_{+i}a+\sigma_{-i}a^\dagger)
\end{equation}
being $\omega$ the frequency of the radiation mode, $a$ and $a^\dagger$ the
ladder operators, $\Delta$ the separation between the energy levels of the
two level atoms, $g$ the coupling and $\sigma_{3i}$,$\sigma_{+i}$ and $\sigma_{-i}$
the Pauli spin matrices. By introducing the operators $S_z=\frac{1}{2}\sum_{i=1}^N\sigma_{3i}$,
$S_\pm=\sum_{i=1}^N\sigma_{\pm i}$ we can rewrite the above Hamiltomian in the form
\begin{equation}
    H = \omega a^\dagger a + \Delta S_z + g(S_+a + S_-a^\dagger).
\end{equation}
A ``ferromagnetic state'' is characterized by having all the two-level atoms in
their ground state. For the sake of simplicity we put us at the resonance $\Delta=\omega$.
Our aim is to study the Tavis-Cummings model in this situation.
This is accomplished by the so called Holstein-Primakoff transformation. This is
defined by
\begin{eqnarray}
    S_+&=&b^\dagger\left(-2S-b^\dagger b\right)^\frac{1}{2} \\ \nonumber
	S_-&=&\left(-2S-b^\dagger b\right)^\frac{1}{2}b \\ \nonumber
%
	S_z&=&S+b^\dagger b
\end{eqnarray}
being $S=-\frac{N}{2}$ and $b$,$b^+$ bosonic operators. The aim of this transformation
is to obtain a series in the parameter $\frac{1}{S}$, being $|S|\gg 1$, that is
pertinent to our situation. These computations are well-known in
literature \cite{pv,ber} but we report it here for completeness:
\begin{eqnarray}
    H_0&=&-\frac{N\omega}{2}+\omega(a^\dagger a + b^\dagger b)+\sqrt{N}g(a^\dagger b + b^\dagger a) \\ \nonumber
	H_1&=&-\frac{g}{2\sqrt{N}}(a^\dagger b^\dagger bb + ab^\dagger b^\dagger b) 
\end{eqnarray}
and the second and third order corrections are given by
\begin{eqnarray}
    H_2 &=& -\frac{g}{8\sqrt{N^3}}(a^\dagger b^\dagger b b^\dagger bb+ab^\dagger b^\dagger b b^\dagger b) \\ \nonumber
	H_3 &=& -\frac{3g}{48\sqrt{N^5}}(a^\dagger b^\dagger b b^\dagger bb^\dagger bb+
	ab^\dagger b^\dagger b b^\dagger bb^\dagger b) 
\end{eqnarray}
from which an easy rule to obtain higher order terms is realized. So, $n$-th order term 
is given by the multiplicative factor 
$-q_n\frac{g}{N^{n-\frac{1}{2}}}$, being $q_n$ the numerical coefficient of the
corresponding order in the series of $(1-x)^\frac{1}{2}$, and Hamilton operator
is given by $a^\dagger\prod_{k=1}^n(b^\dagger b)^k b + h.c.$. Our approximation
holds until the average number of bosonic excitations described by the operator $b$ is
largely smaller than $|S|$.

The leading order Hamiltonian can be immediately diagonalized by introducing two
bosonic operators $c_1$ and $c_2$ as
\begin{eqnarray}
    c_1 &=& \frac{a+b}{\sqrt{2}} \\ \nonumber
	c_2 &=& \frac{a-b}{\sqrt{2}} 
\end{eqnarray} 
that give
\begin{equation}
    H_0 = (\omega + \sqrt{N}g)c_1^\dagger c_1 + (\omega - \sqrt{N}g)c_2^\dagger c_2 
\end{equation}
i.e. two independent harmonic oscillators and we have omitted the constant. This
means that, at the leading order, we expect the coherence to be preserved \cite{pv}.
In sec.\ref{sec3} we will show that this conclusion can be evaded in some way.
Anyhow, the unitary evolution at the leading order is given by
\begin{equation}
    U_0(t) = \exp\left[-it(\omega + \sqrt{N}g)c_1^\dagger c_1\right] 
	\exp\left[-it(\omega - \sqrt{N}g)c_2^\dagger c_2\right].
\end{equation}
The eigenvalues are given by
\begin{equation}
    \epsilon_{n_1n_2}(\omega,g,N) = n_1(\omega+\sqrt{N}g)+n_2(\omega-\sqrt{N}g).
\end{equation}
The eigenstates are given by
\begin{equation}
\label{eq:eig}
    |n1;n2\rangle = \frac{(c_1^\dagger)^{n_1}}{\sqrt{n_1!}}|0\rangle_+
	                \frac{(c_2^\dagger)^{n_2}}{\sqrt{n_2!}}|0\rangle_- 
\end{equation}
having put $|0\rangle_-$ the state having all the two-level atoms in the ground state
on which the $c_2^\dagger$ operator is acting. This clarifies how the time evolution
happens in the limit of $N$ two-level atoms all in their ground state. The next order
correction is $O\left(\frac{1}{N}\right)$.

For a coherent state of the radiation mode
\begin{equation}
    |\psi(0)\rangle = e^{\alpha a^\dagger - \alpha^* a}|0\rangle\left|-\frac{N}{2}\right\rangle
\end{equation}
we will get at the leading order
\begin{equation}
    |\psi(t)\rangle = |\tilde\alpha e^{-i(\omega+\sqrt{N}g)t}\rangle_+|\tilde\alpha e^{-i(\omega-\sqrt{N}g)t}\rangle_-
\end{equation}
with $\tilde\alpha=\frac{\alpha}{\sqrt{2}}$, being
\begin{equation}
    c_1|\tilde\alpha e^{-i(\omega+\sqrt{N}g)t}\rangle_+=
	\tilde\alpha e^{-i(\omega+\sqrt{N}g)t}|\tilde\alpha e^{-i(\omega+\sqrt{N}g)t}\rangle_+
\end{equation}
and
\begin{equation}
    c_2|\tilde\alpha e^{-i(\omega-\sqrt{N}g)t}\rangle_-=
	\tilde\alpha e^{-i(\omega-\sqrt{N}g)t}|\tilde\alpha e^{-i(\omega-\sqrt{N}g)t}\rangle_-
\end{equation}
coherent states of the two harmonic oscillators. These equations give the useful
result for our aims
\begin{equation}
\label{eq:a}
   a|\psi(t)\rangle = \alpha e^{-i\omega t}\cos(\sqrt{N}gt)|\psi(t)\rangle. 
\end{equation} 
From this equation it is easy to recover the well-known results \cite{pv}
\begin{equation}
    \langle \hat n\rangle=\langle\psi(t)|a^\dagger a|\psi(t)\rangle = |\alpha|^2 \cos^2(\sqrt{N}gt)
\end{equation}
being $\hat n$ the number operator for the radiation field and
\begin{equation}
    \langle {\hat n}^2\rangle -\langle \hat n\rangle^2 = |\alpha|^2 \cos^2(\sqrt{N}gt)
\end{equation}
that proves that at the leading order the coherence is kept. This results holds,
at this order, also in the thermodynamic limit $N\rightarrow\infty$, $g\rightarrow 0$
and $\sqrt{N}g = constant$.

\section{Higher Order Corrections to the Unitary Evolution}
\label{sec2b}

The answer to the question about the coherence being kept to higher order has been
properly answered in Ref.\cite{ber}. The answer is given by computing the first order
correction to the eigenvalues by the Rayleigh-Schr\"odinger equation. The Hamiltonian
to use is given by $H_1$ that is, using the operators $c_1$ and $c_2$
\begin{equation}
    H_1=-\frac{g}{4\sqrt{N}}[(c_1^\dagger c_1)^2-(c_2^\dagger c_2)^2+
	c_2^\dagger c_2 -c_1^\dagger c_1 - (c_1^\dagger c_1-c_2^\dagger c_2)(c_1^\dagger c_2 + c_1 c_2^\dagger)
	+c_1^\dagger c_2 - c_2^\dagger c_1]
\end{equation}
and we obtain the correction \cite{ber}
\begin{equation}
    \epsilon^1_{n_1n_2}(g,N) = -\frac{g}{4\sqrt{N}}(n_1^2-n_2^2+n_2-n_1).  
\end{equation}
that gives the following correction to the average of the number operator
\begin{equation}
    \langle \hat n\rangle = |\alpha|^2\sum_{n_1,n_2}e^{-2|\tilde\alpha|^2}
	\frac{|\tilde\alpha|^{2n_1}}{n_1!}\frac{|\tilde\alpha|^{2n_2}}{n_2!}
	\cos\left[\sqrt{N}gt+\frac{g}{4\sqrt{N}}(n_1+n_2)t)\right]
\end{equation}
and we lose coherence if the second term in the argument of the cosine is
comparable to the first one but, in the thermodynamic limit as defined above,
we are granted that coherence is maintained.

It is interesting to compute also the correction to the eigenstates (\ref{eq:eig}).
Rayleigh-Schr\"odinger series gives
\begin{equation}
    |n_1;n_2\rangle_1 = \frac{1}{8N}(n_1\sqrt{n_1}\sqrt{n_2+1}|n_1-1;n_2+1\rangle
	+n_2\sqrt{n_2}\sqrt{n_1+1}|n_1+1;n_2-1\rangle)
\end{equation}
that is $O\left(\frac{1}{N}\right)$ while, as seen above, the correction
to the eigenvalues is $O\left(\frac{1}{\sqrt{N}}\right)$. Then, we are able to
compute the correction to the average of the number operator to first order.
This gives
\begin{eqnarray}
    \delta\langle\hat n\rangle &=& \frac{1}{8N}
	\sum_{n_1,n_2}e^{-2|\hat\alpha|^2}\frac{|\tilde\alpha|^{2n_1}}{n_1!}\frac{|\tilde\alpha|^{2n_2}}{n_2!}
	\left\{(n_1+n_2)(n_1^2+m_1^2)+n_1^2(n_2+1)+n_2^2(n_1+1)\right. \\ \nonumber
	&+&\{[(n_1-1)^2+(n_2+1)^2]\sqrt{n_1}\sqrt{n_2+1}
	+[(n_1+1)^2+(n_2-1)^2]\sqrt{n_1+1}\sqrt{n_2}\}\times \\ \nonumber
    & &\cos\left[2\sqrt{N}gt+\frac{g}{2\sqrt{N}}(n_1+n_2-1)t)\right] \\ \nonumber
	&+&[(n_2+2)\sqrt{n_1}\sqrt{n_2+1}\sqrt{n_1-1}\sqrt{n_2+2}
	+(n_1+2)\sqrt{n_2}\sqrt{n_1+1}\sqrt{n_2-1}\sqrt{n_1+2}]\times \\ \nonumber
    & &\left.\cos\left[4\sqrt{N}gt+\frac{g}{\sqrt{N}}(n_1+n_2-1)t)\right]
	\right\}
\end{eqnarray}
and again we have a confirmation that the thermodynamic limit grants that
coherence is kept for an initial coherent state. As our aim is to apply the time
evolution to a superposition of coherent states, this conclusion is crucial for 
our result to hold in the thermodynamic limit.

\section{Unitary Evolution of a Macroscopic Superposition State}
\label{sec3}

Now, let us consider a phase Schr\"odinger cat state \cite{sch}
\begin{equation}
    |\psi_S(0)\rangle = {\cal N}(|\gamma e^{i\phi}\rangle + |\gamma e^{-i\phi}\rangle)
\end{equation}
being 
\begin{equation}
    {\cal N}^2=\frac{1}{2}\frac{1}{1+\cos(\gamma^2\sin(2\phi))e^{-\Delta^2}}, 
\end{equation}
$\Delta^2=2\gamma^2\sin^2\phi$ the distance between the coherent states, 
$\gamma$ and $\phi$ two real numbers. The time evolution of this state is given by
\begin{equation}
    |\psi_S(t)\rangle = {\cal N}\left[
	|\tilde\gamma e^{-i[(\omega+\sqrt{N}g)t-\phi]}\rangle_+
	|\tilde\gamma e^{-i[(\omega-\sqrt{N}g)t-\phi]}\rangle_-
	+
	|\tilde\gamma e^{-i[(\omega+\sqrt{N}g)t+\phi]}\rangle_+
	|\tilde\gamma e^{-i[(\omega-\sqrt{N}g)t+\phi]}\rangle_-
	\right]
\end{equation}
with $\tilde\gamma=\frac{\gamma}{\sqrt{2}}$.
The computation of the averages is rather straightforward and gives using eq.(\ref{eq:a})
\begin{equation}
    \langle\psi_S(t)|a^\dagger a|\psi_S(t)\rangle = \gamma^2
	\cos^2(\sqrt{N}gt)\frac{1+\cos(2\phi+\gamma^2\sin(2\phi))e^{-\Delta^2}}
	{1+\cos(\gamma^2\sin(2\phi))e^{-\Delta^2}}
\end{equation}
that has the property to give the result of a single coherent state or $\phi\rightarrow 0$ but,
mostly important, for $\Delta\gg 1$ one recover the same result of a single coherent state. So,
more distant are the coherent states and easier the superposition is removed in the time
evolution. This is confirmed by the computation:
\begin{eqnarray}
   \langle\psi_S(t)|a^\dagger aa^\dagger a|\psi_S(t)\rangle&=&\gamma^4
	\cos^4(\sqrt{N}gt)\frac{1+\cos(4\phi+\gamma^2\sin(2\phi))e^{-\Delta^2}}
	{1+\cos(\gamma^2\sin(2\phi))e^{-\Delta^2}} \\ \nonumber
	&+&\gamma^2
	\cos^2(\sqrt{N}gt)\frac{1+\cos(2\phi+\gamma^2\sin(2\phi))e^{-\Delta^2}}
	{1+\cos(\gamma^2\sin(2\phi))e^{-\Delta^2}}
\end{eqnarray}
that gives back the result of a single coherent state both for $\phi=0$ and
$\Delta\gg 1$ confirming that the unitary evolution for the Tavis-Cummings model,
in the thermodynamic limit, grants the disappearance of the superposition state.
This situation has been encountered also for the Dicke model in Ref.\cite{fra3}
where it was shown an identical result in this case but resorting to the
concept of singular limits.
Here, the argument still relies on the thermodynamic limit, that is essential to the proof, 
but we have also assumed the components of the Schr\"odinger cat state are enough distant, $\Delta\gg 1$.

It is important to note that, if the coherent states in the superposition are
truly macroscopic, we are left with a single coherent state behaving as a classical
radiation field.  

\section{Conclusions}
\label{sec4}

In conclusion, we have shown how decoherence can be produced in the thermodynamic limit also for
the N atom Jaynes-Cummings model that can be straightforwardly used to test the very
existence of a somewhat different kind of decoherence, even if the N atoms can be seen
as an environment. But decoherence appears dynamically.

We would like to emphasize the duality of our approach with respect to the one
of Gea-Banacloche \cite{gb}.
Striking evidence of classicality emerging by increasing the number of photons in a cavity,
without resorting to any concept of environment, has been given recently in the experiment
of Haroche's group \cite{har2}, giving full support to the analysis of Gea-Banacloche \cite{gb}. 
So, it would be interesting to see also 
states for $N$ atoms in the thermodynamic limit as obtained here. Then, decoherence
can also appear as an intrinsic effect of the unitary evolution in the
thermodynamic limit. 


\end{document}